\documentclass[UTF8,a4paper]{article}
\usepackage{amsmath}
\usepackage{graphicx}
\usepackage{subfigure}
\usepackage{epstopdf}
\usepackage{geometry}
\usepackage{ulem}
\usepackage{indentfirst}
\usepackage{amsfonts,amssymb,dsfont}
\usepackage{setspace}
\usepackage{threeparttable}
\usepackage{amsmath,mathtools,amsthm}
\usepackage{array}
\usepackage{extarrows}
\usepackage{upgreek}

\makeatletter
\renewcommand*\env@matrix[1][\arraystretch]{%
  \edef\arraystretch{#1}%
  \hskip -\arraycolsep
  \let\@ifnextchar\new@ifnextchar
  \array{*\c@MaxMatrixCols c}}
\makeatother
\begin{document}


\title{\bf Many-electron effect to the dynamical polarization of silicene-like two-dimension Dirac materials}
\author{Chen-Huan Wu
\thanks{chenhuanwu1@gmail.com}
\\Key Laboratory of Atomic $\&$ Molecular Physics and Functional Materials of Gansu Province,
\\College of Physics and Electronic Engineering, Northwest Normal University, Lanzhou 730070, China}

\maketitle
\vspace{-30pt}
\begin{abstract}
\begin{large} 

We discuss the dynamical polarization with finite momentum and frequency in the presence of many-electron effect,
including the screened Coulomb interaction, self-energy and vertex correction.
The longitudinal conductivity, screened Coulomb interaction, and the response function are calculated.
The behavior of the Dirac Fermions, including the propagation of the charge density which exhibits the causality,
affects largely the low-temperature physical properties of the Dirac semimetal, like the silicene.
For the polarization-related quantities (like the dielectric function), the method of standard random phase approximation (RPA) 
provides the non-interaction results (ignore the many electron effect),
for a more exact result, we discuss the self-energy and the vertex correction for the two-dimension Dirac model.
We found that,
after the self-energy correction, the longitudinal conductivity increase compared to the noninteracting one in optical limit.
For the renormalization treatment, the ultraviolet cutoff is setted as $\Lambda=t$ in our calculations,
i.e., within the range between two Van Hove singularities where the density of states divergent logarithmically.
The (corrected) screened Coulomb interaction and the response function are also discussed.
Our results are helpful to the application of the Dirac materials (or the Weyl semimetal) in spintronics or valleytronics.\\



\end{large}

\end{abstract}
\begin{large}
\section{Introduction}

As it's well known that the dynamical polarization\cite{Wu C H3,XX} is important to explore the two-dimension materials including the two-dimension electron gas (2DEG)
in the presence of finite scattering wave vector ${\bf q}$ and frequency $\omega$ (like the Matsubara frequency of the Fermions).
It's also important to understand the properties like the screening of the Coulomb potential by the charged impurities, and the dielectric function 
as well as the optical absorption when under a light field with certain frequency\cite{XX}.
The many-body effect (many-electron effect) is also related to the retarded polarization function as well as the dielectric function,
thus in the following computation, we define $\omega=\omega+i0^{+}$ for simplicity.
The self-energy effect in the many-electron system is generally important for the system with quasi-one-dimension band,
like the graphene, silicene, and genmanene in the $M-K$ region\cite{Wu C H1,Wu C H2,Wu C H3,Wu C H4,Wu C H5,Wu C H7,Wu C HX} 
or the black phosphorus in the $M-Y$ region\cite{Tran V}.
The ignorance of the electron-interaction effect also leads to the ignorance of the self-energy and the related excitonic effect
and the Coulomb potential.
The ignorance of self-energy will cause the underestimation of the band gap and upvaluation of the optical-absorption
as we discuss below.
More importantly, the method of random phase approximation (RPA) with the interaction correction may underestimates the dielectric function
and overestimates the dynamical polarization in the static-limit\cite{Sodemann I}.

Due to the interaction correction, the charge-density-dependence (of the density of states and plasmon frequency) is rised
together with the self-consistently determined screened impurity potential.
For the polarization-related quantities (like the dielectric function), the RPA
provides the non-interaction result (except the long-range divergence of the Coulomb interaction\cite{Sensarma R})
i.e., it underestimates and overestimates the dielectric function for $\omega<v_{F}{\bf q}$ and $\omega>v_{F}{\bf q}$, respectively\cite{Sodemann I},
and thus underestimates and overestimates the polarization for $\omega>v_{F}{\bf q}$ and $\omega<v_{F}{\bf q}$, respectively.
For the many-body effect to the dielectric function, 
the method of scanning tunneling microscopy (STM) can provide a more exact result.
Through the lattice Green function (where we compute in the zero temperature limit),
the local-field effect (the Hubbard term),
which being ignored in the optical limit of the dielectric function, 
can also be taken into account\cite{Sensarma R}.
In this article, we also discuss the plasmon model dispersions within RPA and the long-wavelength limit,
the self-energy correction can also be taken into account by the corrected Fermi velocity (phase velocity)
which has been experimentally observed\cite{Elias D C}.
While the plasmon model in three dimension Weyl semimetal has been discussed in Refs.\cite{Lv M,Zhou J}.
Due to the corrected velocity, the linear Dirac spectrum is replaced by the quadratic one in the low-energy regime (small ${\bf q}$).

\section{Model}

We at first intruduce the
low-energy Dirac-Hamilatonian of silicene in tight-binding model,
which reads\cite{Wu C HX,Wu C H3,Wu C H1,Wu C H4,Wu C H7,XX}
\begin{equation} 
\begin{aligned}
H=&\hbar v_{F}(\eta\tau_{x}k_{x}+\tau_{y}k_{y})+\eta\lambda_{{\rm SOC}}\tau_{z}\sigma_{z}+a\lambda_{R_{2}}\eta\tau_{z}(k_{y}\sigma_{x}-k_{x}\sigma_{y})\\
&-\frac{\overline{\Delta}}{2}E_{\perp}\tau_{z}+\frac{\lambda_{R_{1}}}{2}(\eta\sigma_{y}\tau_{x}-\sigma_{x}\tau_{y})+M_{s}s_{z}+M_{c}
-\eta\tau_{z}\hbar v_{F}^{2}\frac{\mathcal{A}}{\Omega}+\mu,
\end{aligned}
\end{equation}
where 
$E_{\perp}$ is the perpendicularly applied electric field, 
$a=3.86$ is the lattice constant,
$\mu$ is the chemical potential,
$\overline{\Delta}=0.46$ \AA\ is the buckled distance between the upper sublattice and lower sublattice,
$\sigma_{z}$ and $\tau_{z}$ are the spin and sublattice (pseudospin) degrees of freedom, respectively.
$\eta=\pm 1$ for K and K' valley, respectively.
$M_{s}$ is the spin-dependent exchange field and $M_{c}$ is the charge-dependent exchange field.
$\lambda_{SOC}=3.9$ meV is the strength of intrinsic spin-orbit coupling (SOC) and $\lambda_{R_{2}}=0.7$ meV is the intrinsic Rashba coupling
which is a next-nearest-neightbor (NNN) hopping term and breaks the lattice inversion symmetry.
$\lambda_{R_{1}}$ is the electric field-induced nearest-neighbor (NN) Rashba coupling which has been found that linear with the applied electric field
in our previous works\cite{Wu C H1,Wu C H2,Wu C H3,Wu C H4,Wu C H5}, which as $\lambda_{R_{1}}=0.012E_{\perp}$.
We here ignore the effects of the high-energy bands on the low-energy bands.
The Dirac-mass and the corresponding quasienergy spectrum (obtained throught the diagonalization procedure) are\cite{Wu C HX}
\begin{equation} 
\begin{aligned}
&m_{D}=\eta\sqrt{\lambda_{{\rm SOC}}^{2}+a^{2}\lambda^{2}_{R_{2}}k^{2}}s_{z}\tau_{z}-\frac{\overline{\Delta}}{2}E_{\perp}\tau_{z}+M_{s}s_{z},\\
&\varepsilon=s\sqrt{a^{2}\lambda^{2}_{R_{2}}k^{2}+(\sqrt{\hbar^{2}v_{F}^{2}{\bf k}^{2}
+(\eta\lambda_{{\rm SOC}}s_{z}\tau_{z}-\frac{\overline{\Delta}}{2}E_{\perp}\tau_{z} )^{2}}+M_{s}s_{z}+s\mu)^{2}},
\end{aligned}
\end{equation}
respectively, 
$s=\pm 1$ is the electron/hole index.
The Dirac-mass here is related to the band gap by the relation $\Delta=2|m_{D}|$.

\section{interaction correction due to the many-electron effect}

To take the many-electron effect (the interaction correction) into consideration,
we firstly need to introduce the
dimensionless nonlocal
Coulomb coupling parameter $g=\frac{e^{2}}{\epsilon_{0}\epsilon\hbar v_{F}}\approx\frac{2.16}{\epsilon_{0}\epsilon}$\cite{Khveshchenko D V2}
where
$\epsilon=(3.9+1)/2=2.45$ is the static background dielectric constant for the air/SiO$_{2}$ surrounding medium of the silicene,
and it's found that also fits the result of the bulk graphite in optical limit (long wavelength-limit)\cite{Wehling T O}.
This Coulomb coupling parameter is also much larger than the fine-structure constant.
In another generate form, it can also be 
represented by the ratio between the Coulomb potential and the thermodynamic potential:
\begin{equation} 
\begin{aligned}
g&=\frac{V_{C}({\bf r})}{V_{\beta}}
=\frac{V({\bf r})}{(-\frac{4}{2\pi\hbar^{2}v_{F}^{2}}\sum_{\mu}[\int\frac{1}{1+e^{\beta(E-\mu)}}d\mu-\mu])}\\
&=\frac{(e^{2}/(4\pi\epsilon|{\bf r}|))}{(-\frac{4}{2\pi\hbar^{2}v_{F}^{2}\beta}\sum_{\mu}{\rm ln}(1+e^{\beta(E-\mu)})}
\end{aligned}
\end{equation}
where the parameter $g_{s}g_{v}=4$ denotes the spin and valley degenerate
as we assumed here to sum up all the Fermion species.

In the presence of finite Dirac-mass, 
the longitudinal conductivity reads
\begin{equation} 
\begin{aligned}
\sigma({\bf q},\omega)=\frac{i\omega e^{2}}{{\bf q}^{2}}\Pi({\bf q},\omega).
\end{aligned}
\end{equation}
In the optical limit (${\bf q}\rightarrow 0$) where ${\bf q}$ is the scattering wave vector by the charged impurities
or the polarized vector in the direction identical with the polarized light, the above longitudinal conductivity reads
\begin{equation} 
\begin{aligned}
\sigma({\bf q},\omega)
=\lim_{{\bf q}\rightarrow 0}\frac{i\omega e^{2}}{{\bf q}^{2}}\frac{2e^{2}\mu g_{s}g_{v}}{\epsilon_{0}\epsilon\hbar^{2}v_{F}^{2}}.
\end{aligned}
\end{equation}
For the noninteracting case (or with the screened Coulomb potential of the charged impurity),
we present in Fig.1 the results of the dynamical polarization and longitudinal conductivity in the cases of $\mu>m_{D}^{\eta\sigma_{z}\tau_{z}}$ and 
$\mu<m_{D}^{\eta\sigma_{z}\tau_{z}}$.
In the real part of the dynamical polarization,
the kink in ${\bf q}=2{\bf k}_{F}$ can be easily found (see also Ref.\cite{XX}),
which with nonanalytic singularities in the first derivative of polarization of the two-dimension Dirac materials (in gapped case)
or the three-dimension trivial materials,
and in the second derivative of polarization of the two-dimension Dirac materials (in gapless case) or the Weyl semimetals\cite{Lv M}.
The singularities in these points also leads to the Friedel oscillation of the screened potential,
which is much stronger than the one observed in the decay of the density of state due to the scattering\cite{Feng B}.

We also see that, in the case of $\mu>m_{D}^{\eta\sigma_{z}\tau_{z}}$, the longitudinal conductivity is zero in the optical limit 
(and independent of the photon energy $\omega$).
While in the case of $\mu<m_{D}^{\eta\sigma_{z}\tau_{z}}$,
the longitudinal conductivity is still zero in the intraband single particle excitation regime (i.e., the regime with $\omega<v_{F}{\bf q}$),
but nonzero in the interband single particle excitation regime.
Here we note that the sign of the chemical potential unaffects the results of the dynamical polarization
as verified in Ref.\cite{Zhou J,Wan X,Pyatkovskiy P K}.
In the case of zero band gap and electron-hole symmetry with $\mu=0$,
the longitudinal conductivity can be rewritten as
\begin{equation} 
\begin{aligned}
\sigma({\bf q},\omega)
=\lim_{{\bf q}\rightarrow 0}g_{s}g_{v}\frac{i\omega e^{2}}{{\bf q}^{2}}\frac{e^{2}{\bf q}^{2}}{4\epsilon_{0}\epsilon\sqrt{\hbar^{2}v_{F}^{2}{\bf q}^{2}-\hbar^{2}\omega^{2}}}
[i{\rm acosh}(\frac{\omega}{v_{F}{\bf q}})-i{\rm acosh}(-\frac{\omega}{v_{F}{\bf q}})].
\end{aligned}
\end{equation}
Fig.2 shows the longitudinal conductivity in optical limit in the presence of background dielectric constant $\epsilon_{0}\epsilon=2.45$.
We obtained a longitudinal conductivity smaller than the ideal frequency-independent result $e^{2}/4\hbar=0.25$
(we estimate $\hbar=e=1$ here) which is predicted by Ref.\cite{Ludwig A W W}.

Under the effect of surface reconstruction,
the Fermi velocity obtained by renormalization group theory is\cite{Sodemann I}
\begin{equation} 
\begin{aligned}
v_{F}'=v_{F}+\frac{e^{2}}{4\epsilon_{0}\epsilon}{\rm ln}\frac{\Lambda}{{\bf q}}
\end{aligned}
\end{equation}
where the high-momentum cutoff $\Lambda$ here is estimated equal to the intrinsic nearest-neighbor hopping $t=1.6$ eV,
i.e., the location of the Van Hove singularities in momentum space.
The effect of the states within the energy window between $\Lambda$ and the bandwidth $W\approx 3t=4.8$ eV
are taken into consideration within the RPA procedure,
but for the logarithmic correction of the dynamical polarization,
the states from zero to the $\Lambda$ are integrated due to the logarithmic divergent of the density of states and 
the $\Lambda$ can be estimated as a free parameter (nonuniversal) much larger than the ${\bf q}$ and $\omega$ as shown in the procedure\cite{Zhou J,Sensarma R,Lv M}.
Sometimes the $\Lambda$ also estimated as 1 for simplicity as done in Ref.\cite{Sensarma R}.
Here we note that the density of states of Fermi level also related to the charge density $n$,
which acts as $D_{F}=\sqrt{g_{s}g_{v}\frac{n}{\pi}}\frac{1}{\gamma}\approx n^{1/2}$ for the two-dimension Dirac system,
and acts as $D_{F}=^{3}\sqrt{g_{s}g_{v}\frac{9n^{2}}{2\pi^{2}}}\frac{1}{\gamma}$ for the three-dimension Dirac or Weyl system,
where $\gamma$ is the band parameter which $\sim v_{F}$ in the case of large charge density\cite{Wu C H3}.
Similar charge density-dependence can be found in the plasmon frequency,
which acts as $\omega_{p}\approx n^{1/2}$ for bilayer silicene or the 2DEG,
and acts as $\omega_{p}\approx n^{1/4}$ for monolayer silicene\cite{Sensarma R}.

We can see that the velocity is enlarged by the effect of self-energy (in the absence of scattering of the charged impurity)
in small ${\bf q}$ region where ${\bf q}<\Lambda$,
that's in contrast to the effect of the on-site Hubbard interaction\cite{Jafari S A}
as we investigated before\cite{Wu C H1}.
That results in the decrease of the Coulomb coupling,
and in weak Coulomb coupling limit $g\ll 1$, the momentum-dependence is logarithmically decrease while the frequency-dependence is increase.
Due to this, the results after the self-energy correction are also independent of the value of ultraviolet cutoff $\Lambda$ in the optical limit
as we discuss below.

Due to the above renormalized Fermi velocity,
the absorption of the radiation in optical limit is also been modified, which reads
\begin{equation} 
\begin{aligned}
A(0,\omega)=\frac{\pi\alpha_{0}}{n_{t}}(1+\frac{\hbar^{2}\omega^{2}}{16t^{2}}),
\end{aligned}
\end{equation}
where $\alpha_{0}=\frac{e^{2}}{2\epsilon_{0}h c}=1/137.036$ is the Sommerfeld fine structure constant,
$n_{t}$ is the relative refractive index which can be estimated as 1 here,
$t$ is the nearest-neighbor hopping modified by the light field and spreads out over a range of $0.4$ eV $\sim$ $1.09$ eV
in the absence of interction correction\cite{Matthes L}
due to the perturbation from small $\omega$ ($<$2 eV) and it's
$t=1.6$ eV for the $\omega=0$ case.
In the $\omega\rightarrow 0$ limit, the university absorption of silicene can be obtained as $\alpha(0,0)=0.0229$\cite{Matthes L,Mak K F}.
That's also the absorption of many other two-dimension materials of the group-IV,
but it's invalid for the three-dimension bulk materials which with a momentum-dependent integral on the constant-energy surface\cite{Huang R}
around the Dirac-point.
Here we note that, a zero-refractive-index phenomenon (and thus with zero dielectric function) can be found in the gapless Dirac-cone\cite{Huang X,Suchowski H}
which with a infinite phase velocity of light.
In such low-frequency limit, the effect of the interaction correction is very small and hard to observed in experiments.
Through the above renormalized Fermi velocity,
the corrected optical absorption in small $\omega$ regime can be obtained.

We present in Fig.3 the optical absorption after interaction correction. 
We can clearly see that the linear Dirac-spectrum is missing even in the low-energy.
Due to the existence of the self-energy,
the optical absorption which is begin from $\alpha_{0}\pi$ in the absence of SOC has a lower rate of increase compared to the one in the absence of
self-energy. The density function theory (DFT) calculation result is also shown by the crosses.
Note that here we carry out the DFT calculation base on the Quantum-ESPRESSO package\cite{Giannozzi P}, and
the wave energy
cutoff is setted as 400 eV for the ultrasoft pseudopotential in our calculations.

\section{Coulomb interaction and response function}

The high-energy state can screen the Coulomb potential which with the in-plane long-range Hubbard repulsion.
Such high-energy state can be found in the Van Hove-singularities at the $M$-point of the Brillouin zone boundary,
or the high-energy $\pi$-plasmon which with high high peak (about 5 eV) in the low-$\omega$ region of the energy loss function\cite{XX}.
The Van Hove-singularities with high density also with highest value of the interband transition matrix element.
The screened two-dimension Coulomb potential within RPA reads
\begin{equation} 
\begin{aligned}
V({\bf q},\omega)=\frac{V_{0}}{1-g_{s}g_{v}V_{0}\Pi({\bf q},\omega)},
\end{aligned}
\end{equation}
For bilayer system, the Coulomb potentail can be rewritten as\cite{Baskaran G,Hawrylak P}
\begin{equation} 
\begin{aligned}
V_{bi}({\bf q},\omega)=\frac{V_{0}{\rm sinh}(d{\bf q})}{\sqrt{({\rm cosh}(d{\bf k})+V_{0}{\rm sinh}(d{\bf q})\Pi({\bf q},\omega))^{2}-1}},
\end{aligned}
\end{equation}
where $d\approx 2.5$ \AA\ is the interlayer distance,
$V_{0}=\frac{2\pi e^{2}}{\epsilon_{0}\epsilon{\bf q}}$ is the Fourier transform of two-dimension bare Coulomb interaction (propagator).
For the Bulk structure formed by the multi layers , like the graphite, multi-layer black phosphorus or silicene, the Coulomb potentail reads
\begin{equation} 
\begin{aligned}
V_{bu}({\bf q},\omega)=\frac{V_{0}F(q_{z})d}{1-g_{s}g_{v}V_{0}\Pi({\bf q},\omega)F(q_{z})},
\end{aligned}
\end{equation}
where $F(q_{z})={\rm sinh(d{\bf q})}/[{\rm cosh}(d{\bf q})-{\rm cos}(dq_{z})]$ with $q_{z}$ the out-of-plane component of the scattering vector.

The response function\cite{Baskaran G,Reed J P} for the disturbance can be obtained by the above Couomb potential 
through the relation 
\begin{equation} 
\begin{aligned}
\chi({\bf q},\omega)=V({\bf q},\omega)\frac{\Pi({\bf q},\omega)}{V},
\end{aligned}
\end{equation}
where $V$ is the Coulomb propagator in corresponding dimension.
Then the response function can yields the propagator of the charge density as a causal result, which can be related to the temperature in 
the fluctuation-dissipation theorem, and reads 
$S({\bf q},\omega)=\frac{1}{\pi}\frac{1}{e^{-\omega/k_{B}T}-1}{\rm Im}[\chi({\bf q},\omega)]$\cite{Abbamonte P,Yang L,Doi M,Milner S T}.
The screened Coulomb interaction and the response function are presented in the Fig.5 for three different Dirac-mass.

It's known that the interband electron-hole pair excited from the, e.g., plasmon which is a collective model of the oscillating electrons,
can only exist in the regime of $\omega>v_{F}{\bf q}$.
In this regime, the interband transition in optical limit can also be described by the dynamical structure factor
\begin{equation} 
\begin{aligned}
S(\omega)=\lim_{{\bf q}\rightarrow 0}\sum_{c,v,{\bf q}}
\left|\frac{e\hbar}{i\sqrt{4\pi\epsilon_{0}}m_{0}}\frac{\langle\psi;{\bf k}|m_{0}{\bf v}\cdot{\bf e}_{\bf q}|\psi';{\bf k}\rangle}{\varepsilon_{\psi}-\varepsilon_{\psi'}}\right|^{2}
\delta(\omega-\varepsilon_{c}+\varepsilon_{v}),
\end{aligned}
\end{equation}
which is complex due to the broken of inversion symmetry in silicene.
For a large range of the photon energy, we can find three peak (two small peak and one main peak) in the optical absorption of the silicene.
In fact, the optical absorption is in a similar shape with the energy-loss function (or dielectric-loss function).
As shown in Fig.4, 
where the results for zero on-site Hubbard interaction and nonzero on-site Hubbard interaction are presented,
the two smaller peak in the absorption and energy-loss function can be seen as indicated,
they are locate in 2 eV and 5 eV respectively.
The former one is due to the $\pi$-plasmon and the later one is due to the $\pi+\sigma$ plasmon.
Here the strength of the Hubbard interaction is setted close to the 
nearest-neighbor Coulomb interaction strength\cite{Wehling T O},
and it has been found related to the phase-transition characters of the silicene\cite{Wu C H1}.
However, we also notice that there are a deprssed peak for the absorption in around 5 eV which can't found in the energy-loss function.
In the presence of the local field effect,
this devation can ne explained by the three optical bright excited electron-hole state\cite{Yang L}.
The anisotropic dispersion of $\pi$-plasmon creases a six-fold pattern of the charge density\cite{Reed J P},
which may be related to the hexagonal warping of the graphene due to the high energy.
Moreover, the excitonic resonant effect can be found in the absorption spectrum near the Van Hove singularities
while the multi-photon resonance can be found in the range of negative photon energy (frequency). 
Through the dynamical polarization depicted in the ${\bf q}\sim{\omega}$ space as shown in Fig.1,
the results for the negative frequency can be obtained easily by the idensity which is causal:
\begin{equation} 
\begin{aligned}
&{\rm Re}\Pi({\bf q},-\omega)={\rm Re}\Pi({\bf q},\omega),\\
&{\rm Im}\Pi({\bf q},-\omega)=-{\rm Im}\Pi({\bf q},\omega),
\end{aligned}
\end{equation}
which is devived from the generate result $\Pi({\bf q},-\omega)=(\Pi({\bf q},\omega))^{*}$.
We want to note that, except the above dynamical optical absorption $-{\rm Im}[1/\epsilon(0,\omega)]$,
the $\pi$-plasmon and the $\pi+\sigma$ plasmon cab also be probed by the dynamical electron density (mainly the valence electrons)\cite{Reed J P}
in low-momentum regime
or the X-ray scattering method\cite{Abbamonte P} which is causal
and related to the retarded response function as mentioned above, and it's
valid for both the two-dimension or three dimension materials.

\section{Correction from the self-energy and vertex function}

It's known\cite{Wu C H1} that the vertex correlation function $\Gamma$ is associated with the jumping of the self-energy
with different spectral weight in the single particle excitation spectrum (with the electron-hole pairs),
and it's important for the variant cluster approximation.
In the presence of the impurities scattering, the vertex correlation function arised with the lifting of Fermi level,
which leads to the segmental structure of energy spectrum,
and thus increase the cancellation effect\cite{Grimaldi C} to the Hall conductivity.
Otherwise it can be ignored if Fermi energy is close to zero and thus it's compeletely spin-dependent, as seen in the single-site DMFT\cite{Wu C H1,Wu C H8}.
For multilayer system, although the vertex function is piecewise for a finite Fermi energy,
it can be ignored for the case of $m_{D}=0$ and with zero interlayer hopping through the vertex renormalization\cite{Khveshchenko D V}.

Here we define the unscattering lattice Green's function
\begin{equation} 
\begin{aligned}
G_{0}({\bf q},\omega)=[\varepsilon-H({\bf q})-\Sigma({\bf q},\omega)]^{-1}.
\end{aligned}
\end{equation}
Then the single-site Green's function can be written as
\begin{equation} 
\begin{aligned}
G({\bf q},\omega)=\frac{G_{0}({\bf q},\omega)}{1-\Sigma({\bf q},\omega)G_{0}({\bf q},\omega)},
\end{aligned}
\end{equation}
with
\begin{equation} 
\begin{aligned}
\Sigma({\bf q},\omega)=\int\frac{d^{2}q}{(2\pi)^{2}}G({\bf k}+{\bf q},\omega+\Omega)\frac{V({\bf k},\omega)}{\epsilon({\bf k},\omega)}\Gamma({\bf q},{\bf k},{\bf k}+{\bf q}),
\end{aligned}
\end{equation}
where the three point vertex function $\Gamma({\bf q},{\bf k},{\bf k}+{\bf q})\propto \Sigma({\bf q},\omega)-\Sigma({\bf k},\Omega)$ 
comes from the jump of self-energy 
and in the direction normal to the boundary between different pieces with different segments.
$\frac{V({\bf k},\omega)}{\epsilon({\bf k},\omega)}$ is the screened Coulomb interaction,
where dielectric function can be obtained as
\begin{equation} 
\begin{aligned}
\epsilon({\bf k},\omega)=1-V_{0}\Pi({\bf k},\omega),
\end{aligned}
\end{equation}
as we depicted in the Ref.\cite{XX}.
Here we note that the above single-site Green's function is for the zero-temperature case where the thermal free energy does not taken into accout,
while the detail about the case with finite temperature has presented in our previous work\cite{Wu C H8}.

The dynamical polarization involving the self-energy correction in momentum space reads
\begin{equation} 
\begin{aligned}
\Pi_{{\rm self}}({\bf q},\omega)=\Pi({\bf q},\omega)+2g_{s}g_{v}\int^{\Lambda}_{0}\frac{d^{2}k}{{2\pi}^{2}}\frac{e^{2}k}{4\epsilon_{0}\epsilon}{\rm ln}\frac{\Lambda}{k}
(1-{\rm cos}\theta)
\frac{\hbar^{2}v_{F}^{2}(2k+q)+\omega^{2}}{(\hbar^{2}v_{F}^{2}(2k+q)^{2}-(\omega+i0^{+})^{2})^{2}},
\end{aligned}
\end{equation}
where $\theta=\frac{k+q{\rm cos}\phi}{\sqrt{k^{2}+q^{2}+2kq{\rm cos}\phi}}$ is the angle between ${\bf q}$ and ${\bf k}+{\bf q}$
with $\phi$ the angle between ${\bf k}$ and ${\bf q}$.
The resulting dynamical polarization as well as the longitudinal conductivity in neutral gapless silicene 
are presented in Fig.6.
Here the chemical potential is setted as zero for the invariance of the Fermion density\cite{Lykken J D} (also the current-current correlation)
From Fig.6(b) we can see that the longitudinal conductivity is increased.
But in optical limit, the longitudinal conductivity is still independent of the $\omega$ which is similar to the uncorrected one as mentioned above.

Considering the vertex correction,
the resulting polarization function reads\cite{Sodemann I}
\begin{equation} 
\begin{aligned}
\Pi_{{\rm int}}({\bf q},\omega)=&\Pi_{{\rm self}}({\bf q},\omega)\\
&-g_{s}g_{v}\frac{1}{2}\int^{\Lambda}_{0}\frac{d^{2}k}{(2\pi)^{2}}\int^{k+q}_{|k-1|}\frac{d^{2}k'}{(2\pi)^{2}}
V_{0}(k-k')\\
&\times \frac{\omega^{2}+\hbar^{2} v_{F}^{2}(2k+q)(2k'+q)[(1-{\rm cos}\theta)(1-{\rm cos}\theta')+{\rm sin}\theta{\rm sin}\theta']}
{(\hbar^{2}v_{F}^{2}(2k+q)^{2}-(\omega+i0^{+})^{2})^{2}(\hbar^{2}v_{F}^{2}(2k'+q)^{2}-(\omega+i0^{+})^{2})^{2}},
\end{aligned}
\end{equation}
where $\theta'=\frac{k'+q{\rm cos}\phi}{\sqrt{k'^{2}+q^{2}+2k'q{\rm cos}\phi}}$. 

\section{Plasmon frequency}

In this section, we discuss the plasmon frequency of the doped silicene.
Base on the approximation of zero imaginary part of the polarization, i.e., ${\rm Im}\Pi({\bf q},\omega)=0$,
the plasmon frequency can be approximately obtained by solving the relation ${\rm Re}\Pi({\bf q},\omega)=V_{0}^{-1}$.
Thus base on the equations deduced in the main text, the plasmon frequency (for chemical potential larger than the Dirac-mass) can be obtained,
including the ones after the self-energy correction and vertex correction.
It's found that, in the three-dimension weyl semimetal with chiral anomaly, the metallic nature supports the existence of the plasmon model
\cite{Sarma S D,Zhou J,Zyuzin A A,Lv M}.
The chiral anomaly is due to the electromagnetic response which acts as ${\bf E}\cdot{\bf B}\neq 0$
and thus related to the Berry gauge field (Berry curvature) origin from the magnetic monopole 
and induce two weyl nodes and the chirality-dependent chemical potential.
Except the chiral anomaly, the other anomalous effects in the presence of electromagnetic field as well as the 
Berry curvature have been discussed in Ref.\cite{Wu C HX}.

Through the computation, we found not plasmion model exist in the case of $\mu<m_{D}$,
while for case of $\mu>m_{D}$, where we set $\mu=2$ eV and $m_{D}=0.15$ eV,
the plasmon models are presented in the Fig.7.
In Fig.7, we show the results of the plasmon model obtained by solving:
${\rm Re}\Pi({\bf q},\omega_{p})=V_{0}^{-1}$,
${\rm Re}\Pi({\bf q},\omega_{p})=(g_{s}g_{v})^{-1}V_{0}^{-1}$,
$\Pi({\bf q},\omega_{p}-i\alpha)=(g_{s}g_{v})^{-1}V_{0}^{-1}$,
which correspond to the green circles, red triangles, and blue squares, respectively.
Here $\alpha$ is the decay rate of the plasmon which vanish in the RPA.
The plasmon dispersion of the silicene and 2DEG are also presented in the Fig.7.
We can see that, in small momentum regime, the plasmon dispersions are well agree with the $\sqrt{{\bf q}}$-behavior
which suitable for the low-energy plasmon as we have presented in the Ref.\cite{Wu C H3}.
The long-wavelength result are for the small-${\bf q}$ limit,
but in the order of ${\bf q}=0$, the $\omega_{p}\neq 0$ due to the logarithmic correction\cite{Lv M}. 

Here the regions in ${\bf q}\sim\omega$ space with Dirac-mass $m_{D}=0.15$ eV
can also be observed in graphene or other two-dimension Dirac materials\cite{Wu C H3,XX} as well as the Weyl semimetals\cite{Lv M},
while for the three-dimension trivial materials, the boundary of the single-particle excitation region is determined by the relation 
$\omega=\frac{\hbar^{2}q^{2}}{2m}\pm \hbar v_{F}{\bf q}$.
The quasiparticle chirality of the silicene\cite{Feng B} and the hexagonal warping (with antiparallel spin (or pseudospin))\cite{XX} suppress the 
backscattering of the Dirac Fermions,
and thus reduce the overlap of the spinor,
that may leads to a faster decay of the Friedel oscillation which is $\sim r^{-4}$ (for large $r$) as found in the Weyl semimetal\cite{Lv M},
that can't be found in the graphene.

Base on the our previous results about the real part of the dielectric function (${\rm Re}\varepsilon({\bf q},\omega)$) in Ref.\cite{XX}
where the backgroud dielectric constant is $\epsilon_{0}\epsilon=2.45$,
we conclude that there does not exists the excitonic plasmons under such background dielectric constant 
since the result is positive (${\rm Re}\varepsilon({\bf q},\omega)>0$) in the $\omega< v_{F}{\bf q}$ regime
(see Fig.4 and Fig.7 of Ref.\cite{XX}),
but such excitonic plasmon can be found for the case of $\epsilon_{0}\epsilon=7.3$ as stated in Ref.\cite{Sodemann I}.

\section{Conclusion}

The behavior of the Dirac Fermions, including the propagation of the charge density which exhibits the causality,
affects largely the low-temperature physics (including the pairing instability).
For the polarization-related quantities (like the dielectric function), the method of standard random phase approximation (RPA) only
provides the non-interaction results,
it underestimates and overestimates the dielectric function for $\omega<v_{F}{\bf q}$ and $\omega>v_{F}{\bf q}$, respectively.
The longitudinal conductivity, which independent of the $\omega$ in small ${\bf q}$ limit, also needs an inverse logarithmic correction
(the self-energy correction) in the optical limit,
after the self-energy correction, we find the longitudinal conductivity increase.
In the renormalization treatment, the ultraviolet cutoff is setted as $\Lambda=t$ in our calculations,
i.e., within the range between two Van Hove singularities where the density of states divergent logarithmically.
The $\Lambda$ is indeed a non-universal parameter and can also be choiced larger than arbitary ${\bf q}$ or $\omega$.
The (corrected) screened Coulomb interaction and the response function are also calculated, as well as the optical
absorption of the radiation.
Our results are helpful to the application of the Dirac materials (or Weyl sememetal) in spintronics or valleytronics.

\end{large}
\renewcommand\refname{References}

\clearpage

\clearpage
Fig.1
\begin{figure}[!ht]
   \centering
 \centering
   \begin{center}
     \includegraphics*[width=1\linewidth]{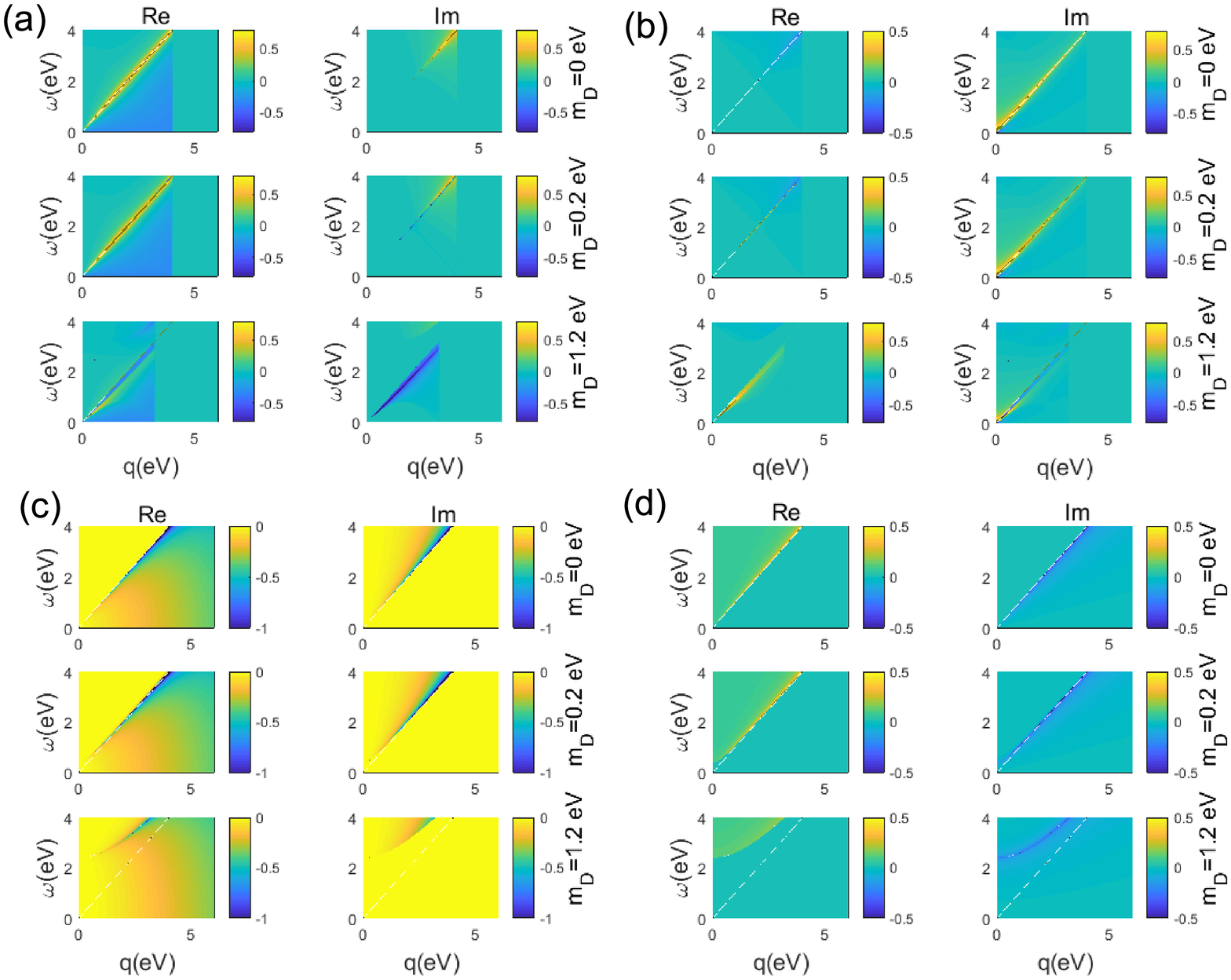}
\caption{(Color online) The dynamical polarization ((a) and (c)) and longitudinal conductivity ((b) and (d)) 
in the cases of $\mu>m_{D}^{\eta\sigma_{z}\tau_{z}}$ ((a) and (b)) and $\mu<m_{D}^{\eta\sigma_{z}\tau_{z}}$ ((c) and (d)).
}
   \end{center}
\end{figure}
\clearpage
Fig.2
\begin{figure}[!ht]
   \centering
 \centering
   \begin{center}
     \includegraphics*[width=0.7\linewidth]{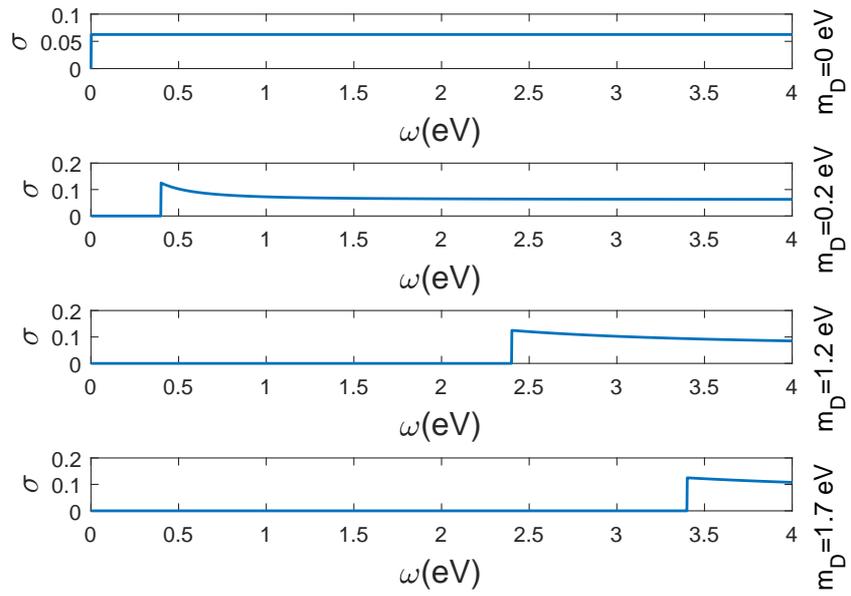}
\caption{(Color online) Longitudinal conductivity $\sigma$ in the case of $\mu<m_{D}$.
}
   \end{center}
\end{figure}
\clearpage
Fig.3
\begin{figure}[!ht]
   \centering
 \centering
   \begin{center}
     \includegraphics*[width=0.7\linewidth]{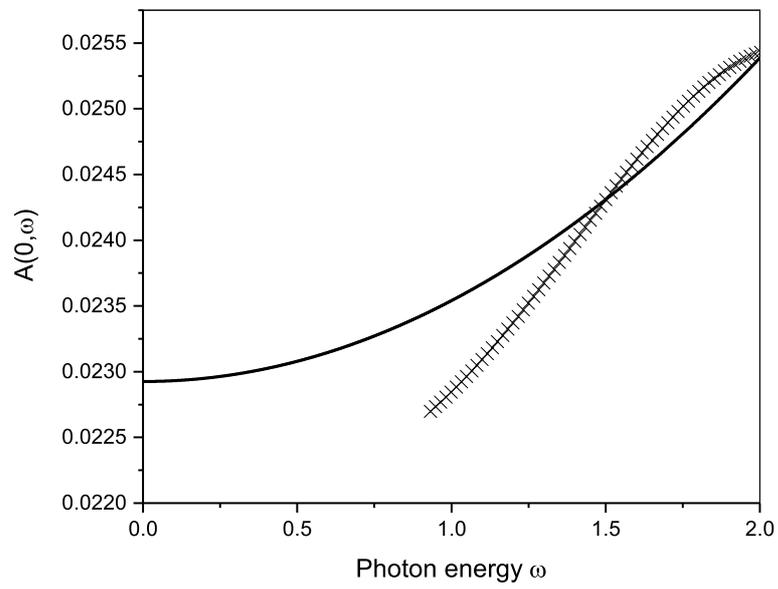}
\caption{(Color online) Optical absorption at small photon energy region (the black line).
The DFT calculation result is indicated by the crosses.
}
   \end{center}
\end{figure}
\clearpage
Fig.4
\begin{figure}[!ht]
   \centering
 \centering
   \begin{center}
     \includegraphics*[width=0.7\linewidth]{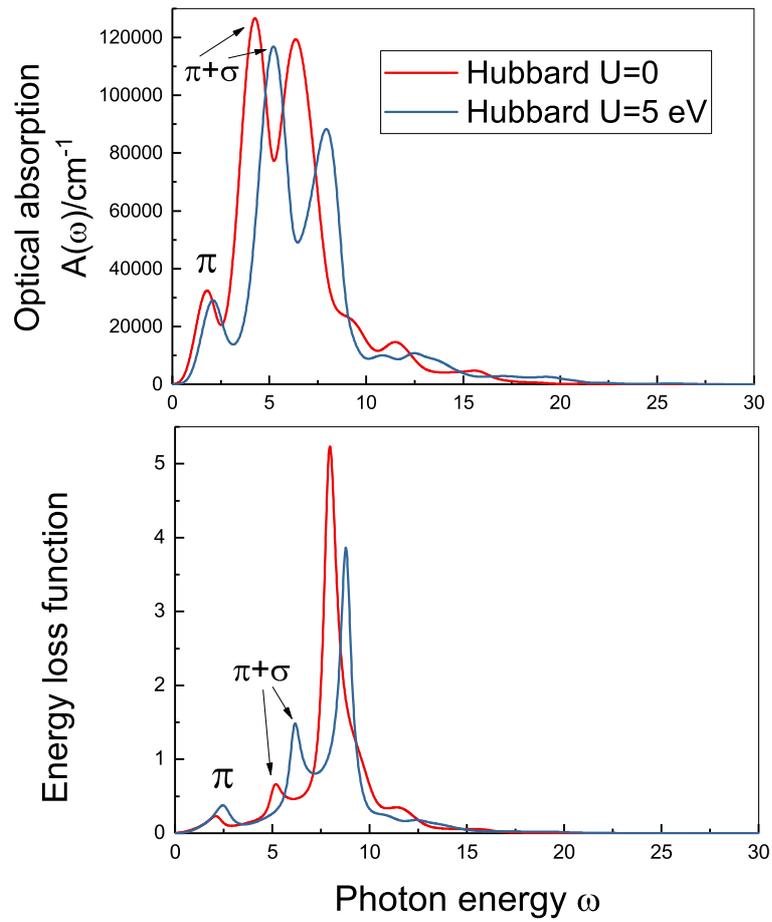}
\caption{(Color online) The optical absorption (a) and energy loss function (b) of silicene with and without Hubbard interaction.
The peaks about the $\pi$-plasmon and $\pi+\sigma$ plasmon are indicated.
}
   \end{center}
\end{figure}
\clearpage
Fig.5
\begin{figure}[!ht]
   \centering
 \centering
   \begin{center}
     \includegraphics*[width=1\linewidth]{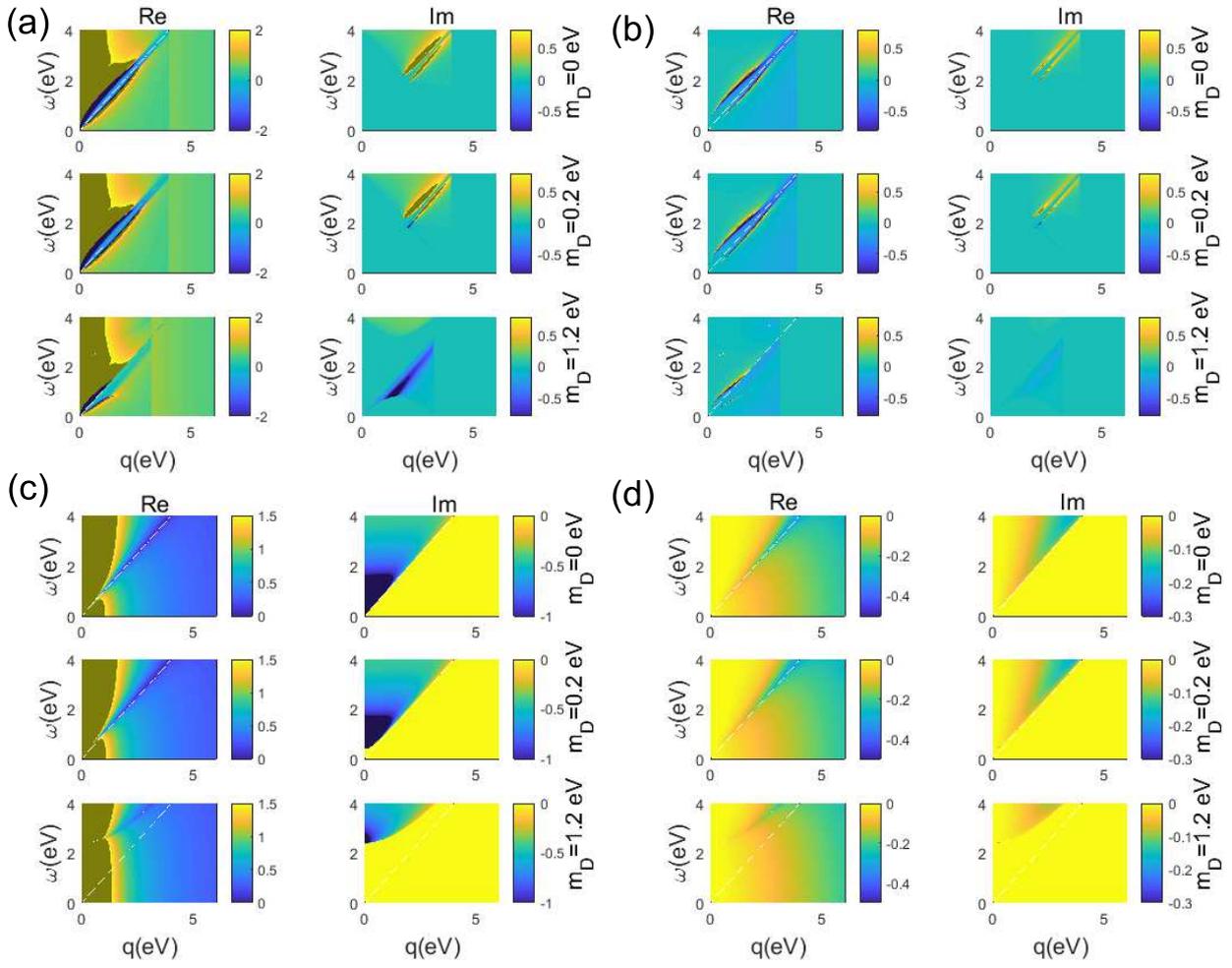}
\caption{(Color online) The screened Coulomb interaction ((a) and (c)) and response function ((b) and (d)) 
in the cases of $\mu>m_{D}^{\eta\sigma_{z}\tau_{z}}$ ((a) and (b)) and $\mu<m_{D}^{\eta\sigma_{z}\tau_{z}}$ ((c) and (d)).
Note that the gray regimes are with very large value (e.g., see Ref.\cite{Wu C H3,Sodemann I,Abbamonte P,Yang L}) including the unphysical behaviors 
which we don't show here.
}
   \end{center}
\end{figure}
\clearpage
Fig.6
\begin{figure}[!ht]
   \centering
 \centering
   \begin{center}
     \includegraphics*[width=0.7\linewidth]{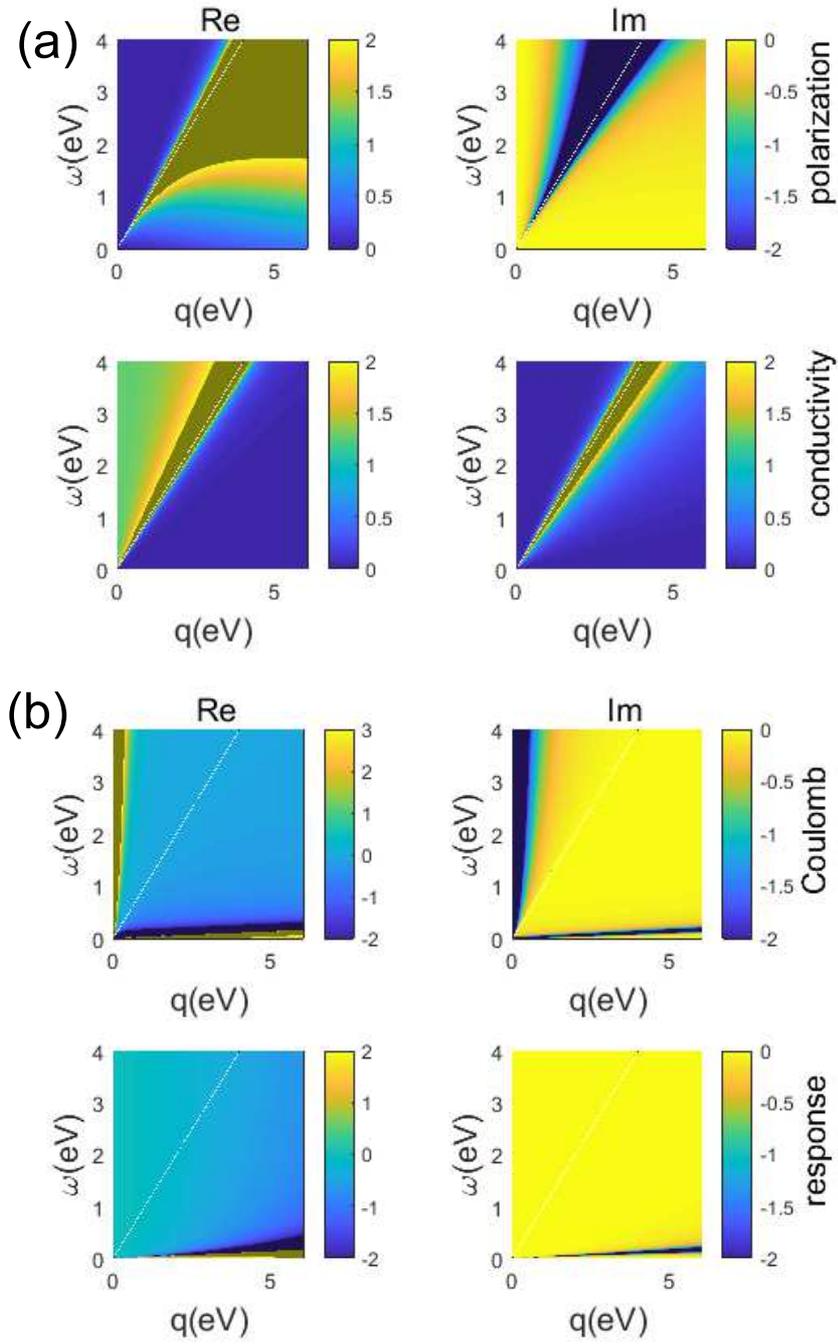}
\caption{(Color online) The dynamical polarization, longitudinal conductivity, Coulomb interaction, and response function (from top to bottom)
after the self-energy correction where we apply the ultraviolent cutoff $\Lambda=t$ here.
}
   \end{center}
\end{figure}
\clearpage
Fig.7
\begin{figure}[!ht]
   \centering
 \centering
   \begin{center}
     \includegraphics*[width=0.7\linewidth]{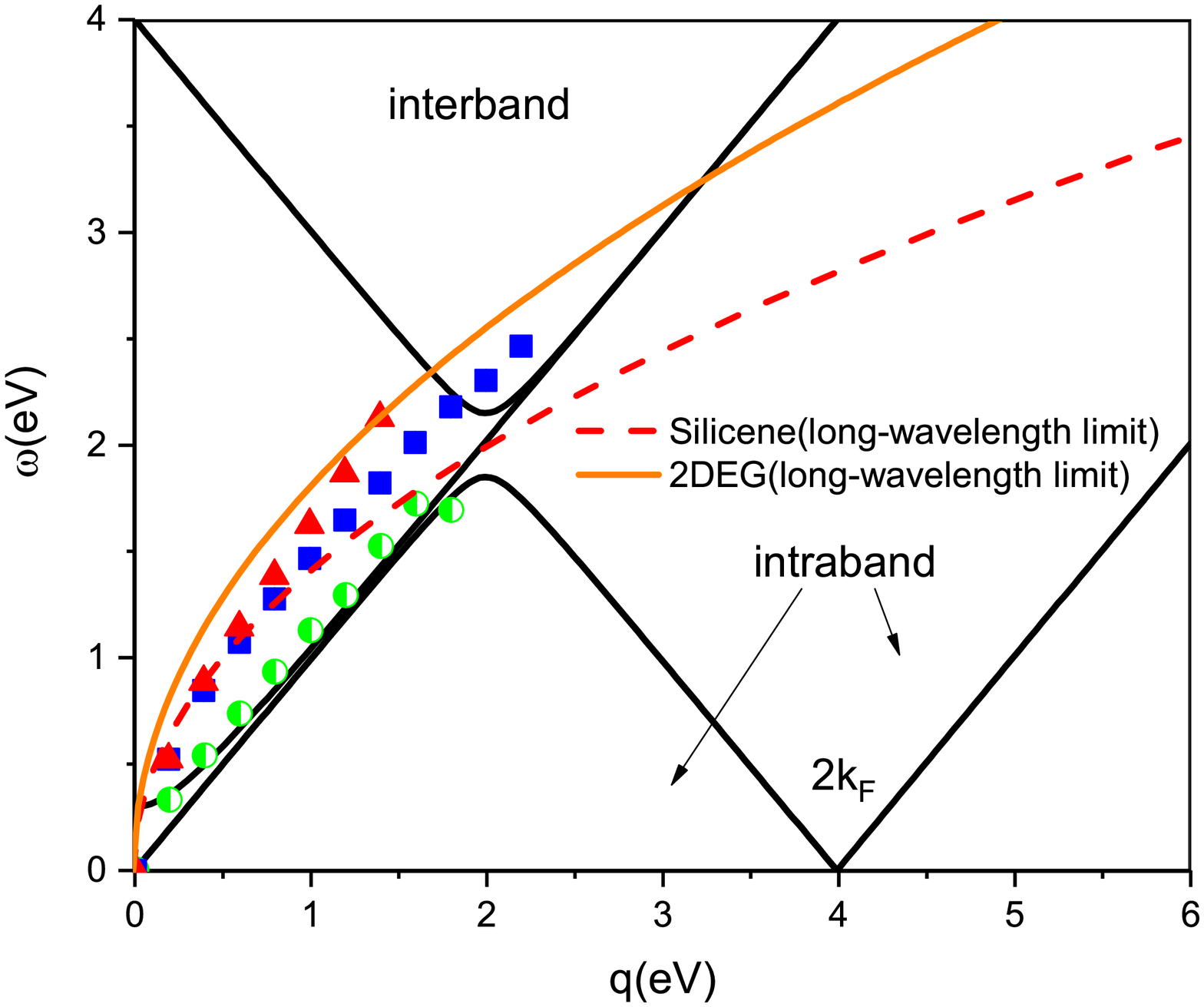}
\caption{(Color online) Plasmon model dispersion in ${\bf q}\sim \omega$ space.
The interband and intraband single-particle excitation regimes are indicated.
The RPA result as well as the long-wavelength result (for the silicene ane 2DEG) are shown
as indicated in the figure.
The Dirac-mass here is setted as $m_{D}=0.15$ eV.
}
   \end{center}
\end{figure}


\begin{thebibliography}{99}
\bibitem{Wu C H3}Wu C H. Interband and intraband transition, dynamical polarization and screening of the monolayer and bilayer silicene in low-energy tight-binding model[J]. arXiv preprint arXiv:1805.07736, 2018.
\bibitem{XX}Wu C H. Dynamical polarization and the optical response of silicene and related materials[J]. arXiv preprint arXiv:1808.03442, 2018.
\bibitem{Wu C H1}Wu C H. Geometrical structure and the electron transport properties of monolayer and bilayer silicene near the semimetal-insulator transition point in tight-binding model[J]. arXiv preprint arXiv:1805.00350, 2018.
\bibitem{Wu C H2}Wu C H. Tight-binding model and ab initio calculation of silicene with strong spin-orbit coupling in low-energy limit[J]. arXiv preprint arXiv:1804.01695, 2018.
\bibitem{Wu C H4}Wu C H. Integer quantum Hall conductivity and longitudinal conductivity in silicene under the electric field and magnetic field[J]. arXiv preprint arXiv:1805.10656, 2018.
\bibitem{Wu C H5}Wu C H. Anomalous Rabi oscillation and related dynamical polarizations under the off-resonance circularly polarized light[J]. arXiv preprint arXiv:1806.03592, 2018.
\bibitem{Wu C H7}Wu C H. Josephson effect in silicene-based SNS Josephson junction: Andreev reflection and free energy[J]. arXiv preprint arXiv:1806.10289, 2018.
\bibitem{Wu C HX}Wu C H. Electronic transport and the related anomalous effects in silicene-like hexagonal lattice[J]. arXiv preprint arXiv:1807.10898, 2018.
\bibitem{Tran V}Tran V, Soklaski R, Liang Y, et al. Layer-controlled band gap and anisotropic excitons in few-layer black phosphorus[J]. Physical Review B, 2014, 89(23): 235319.
\bibitem{Sodemann I}Sodemann I, Fogler M M. Interaction corrections to the polarization function of graphene[J]. Physical Review B, 2012, 86(11): 115408.
\bibitem{Sensarma R}Sensarma R, Hwang E H, Sarma S D. Dynamic screening and low-energy collective modes in bilayer graphene[J]. Physical Review B, 2010, 82(19): 195428.
\bibitem{Elias D C}Elias D C, Gorbachev R V, Mayorov A S, et al. Dirac cones reshaped by interaction effects in suspended graphene[J]. Nature Physics, 2011, 7(9): 701.
\bibitem{Lv M}Lv M, Zhang S C. Dielectric function, Friedel oscillation and plasmons in Weyl semimetals[J]. International Journal of Modern Physics B, 2013, 27(25): 1350177.
\bibitem{Zhou J}Zhou J, Chang H R, Xiao D. Plasmon mode as a detection of the chiral anomaly in Weyl semimetals[J]. Physical Review B, 2015, 91(3): 035114.
\bibitem{Khveshchenko D V2}Khveshchenko D V. Massive Dirac fermions in single-layer graphene[J]. Journal of Physics: Condensed Matter, 2009, 21(7): 075303.

\bibitem{Wehling T O}Wehling T O, {\c{S}}a{\c{s}}{\i}o{\u{g}}lu E, Friedrich C, et al. Strength of effective Coulomb interactions in graphene and graphite[J]. Physical review letters, 2011, 106(23): 236805.
\bibitem{Feng B}Feng B, Li H, Liu C C, et al. Observation of Dirac cone warping and chirality effects in silicene[J]. ACS nano, 2013, 7(10): 9049-9054.
\bibitem{Wan X}Wan X, Turner A M, Vishwanath A, et al. Topological semimetal and Fermi-arc surface states in the electronic structure of pyrochlore iridates[J]. Physical Review B, 2011, 83(20): 205101.
\bibitem{Pyatkovskiy P K}Pyatkovskiy P K. Dynamical polarization, screening, and plasmons in gapped graphene[J]. Journal of Physics: Condensed Matter, 2008, 21(2): 025506.
\bibitem{Ludwig A W W}Ludwig A W W, Fisher M P A, Shankar R, et al. Integer quantum Hall transition: An alternative approach and exact results[J]. Physical Review B, 1994, 50(11): 7526.
\bibitem{Jafari S A}Jafari S A. Dynamical mean field study of the Dirac liquid[J]. The European Physical Journal B, 2009, 68(4): 537-542.
\bibitem{Matthes L}Matthes L, Gori P, Pulci O, et al. Universal infrared absorbance of two-dimensional honeycomb group-IV crystals[J]. Physical Review B, 2013, 87(3): 035438.
\bibitem{Mak K F}Mak K F, Ju L, Wang F, et al. Optical spectroscopy of graphene: from the far infrared to the ultraviolet[J]. Solid State Communications, 2012, 152(15): 1341-1349.
\bibitem{Huang R}Huang R, Li J, Wu Z, et al. Universal absorption of two-dimensional materials within k? p method[J]. Physics Letters A, 2018.
\bibitem{Huang X}Huang X, Lai Y, Hang Z H, et al. Dirac cones induced by accidental degeneracy in photonic crystals and zero-refractive-index materials[J]. Nature materials, 2011, 10(8): 582.
\bibitem{Suchowski H}Suchowski H, O’Brien K, Wong Z J, et al. Phase mismatch–free nonlinear propagation in optical zero-index materials[J]. Science, 2013, 342(6163): 1223-1226.
\bibitem{Giannozzi P}Giannozzi P, Baroni S, Bonini N, et al. QUANTUM ESPRESSO: a modular and open-source software project for quantum simulations of materials[J]. Journal of physics: Condensed matter, 2009, 21(39): 395502.
\bibitem{Baskaran G}Baskaran G, Jafari S A. Gapless spin-1 neutral collective mode branch for graphite[J]. Physical review letters, 2002, 89(1): 016402.
\bibitem{Hawrylak P}Hawrylak P, Eliasson G, Quinn J J. Many-body effects in a layered electron gas[J]. Physical Review B, 1988, 37(17): 10187.
\bibitem{Reed J P}Reed J P, Uchoa B, Joe Y I, et al. The effective fine-structure constant of freestanding graphene measured in graphite[J]. Science, 2010, 330(6005): 805-808.
\bibitem{Abbamonte P}Abbamonte P, Wong G C L, Cahill D G, et al. Ultrafast Imaging and the Phase Problem for Inelastic X‐Ray Scattering[J]. Advanced Materials, 2010, 22(10): 1141-1147.
\bibitem{Yang L}Yang L, Deslippe J, Park C H, et al. Excitonic effects on the optical response of graphene and bilayer graphene[J]. Physical review letters, 2009, 103(18): 186802.
\bibitem{Doi M}Doi M, Shimada T, Okano K. Concentration fluctuation of stiff polymers. II. Dynamical structure factor of rod‐like polymers in the isotropic phase[J]. The Journal of chemical physics, 1988, 88(6): 4070-4075.
\bibitem{Milner S T}Milner S T, Safran S A. Dynamical fluctuations of droplet microemulsions and vesicles[J]. Physical Review A, 1987, 36(9): 4371.
\bibitem{Grimaldi C}Grimaldi C, Cappelluti E, Marsiglio F. Off-Fermi surface cancellation effects in spin-Hall conductivity of a two-dimensional Rashba electron gas. Physical Review B, 2006, 73(8): 081303.
\bibitem{Wu C H8}Wu C H. Time Evolution and Thermodynamics for the Nonequilibrium System in Phase-Space[J]. arXiv preprint arXiv:1711.00547, 2017.
\bibitem{Khveshchenko D V}Khveshchenko D V. Ghost excitonic insulator transition in layered graphite. Physical Review Letters, 2001, 87(24): 246802.

\bibitem{Lykken J D}Lykken J D, Sonnenschein J, Weiss N. Anyonic superconductivity[J]. Physical Review D, 1990, 42(6): 2161.
\bibitem{Sarma S D}Sarma S D, Hwang E H. Collective modes of the massless Dirac plasma[J]. Physical review letters, 2009, 102(20): 206412.
\bibitem{Zyuzin A A}Zyuzin A A, Burkov A A. Topological response in Weyl semimetals and the chiral anomaly[J]. Physical Review B, 2012, 86(11): 115133.






\end{thebibliography}
\end{document}